\documentclass{openjournal}
\def\app#1#2{%
	\mathrel{%
		\setbox0=\hbox{$#1\approx$}%
		\setbox2=\hbox{%
			\rlap{\hbox{$#1\propto$}}%
			\lower1.1\ht0\box0%
			}%
			\raise0.25\ht2\box2%
			}%
			}

\usepackage{revsymb,graphicx,amsmath,amssymb}
\begin{document}
\title{Searching for Periodicity in FRB 20240114A}
\shorttitle{Searching for Periodicity in FRB 20240114A}
\shortauthors{Katz}
\author{J. I. Katz\altaffilmark{}}
\affil{Department of Physics and McDonnell Center for the Space Sciences,
Washington University, St. Louis, Mo. 63130}
\email{katz@wuphys.wustl.edu}
\begin{abstract}
FRB 20240114A is extraordinarily active, and therefore presents an
opportunity to search for the periodicity predicted by magnetar models of
Fast Radio Bursts (FRB).  \cite{Z25} observed 11,553 bursts, including
3196 on MJD 60381 (March 12, 2024).  We find no significant peak in the
periodogram of those 3196 bursts, which occur within 15628 s.  {This
interval is} short enough that even with a characteristic slowing age of 1
year {a periodicity $\ge 0.1\,$s} would not significantly dephase within
the observation.
Introducing modulation artificially shows that an amplitude of 0.15 would
have been detected robustly.
\end{abstract}
\newpage
\section{Introduction}
The most popular model of Fast Radio Bursts (FRB) \citep{Z23} attributes
them to magnetars, hypermagnetized neutron stars.  Magnetars were first
proposed \citep{K82,DT92} as models of Soft Gamma Repeaters (SGR; initially
considered to be gamma-ray bursts), a model that was verified when rapid
spindown between outbursts confirmed their extraordinary magnetic fields.
This model makes one obvious prediction: that FRB, or the level of FRB
activity are periodic with the neutron star's rotational period.  This
prediction has never been verified for any FRB, despite evidence 
\citep{N25,W25} that at least one FRB is inside an expanding supernova
remnant, as would be expected in a magnetar model.

The extreme activity and many detected bursts of FRB 20240114A, including
3196 within 15628 s on the single day March 12, 2024 \citep{Z25}, present a
favorable opportunity for a search.  The 4.3 hour duration of this dataset
is short enough that period derivatives and spin noise \citep{C22} are
unlikely to interfere with detection of a periodicity.  We report the
failure of that search {using the published data of \citet{Z25} for
March 12, 2024.}
\section{Types of periodicity}
We distinguish two types of periodicity:
\subsection{RRAT periodicity}
RRAT (Rotating RAdio Transients) emit pulses whose separations are integer
multiples of a very slowly varying period.  They are interpreted as pulsars
most of whose pulses are nulled.  It was easy to show (\citet{K22} used five
bursts of FRB 20121102 observed within 93 s by \citet{G18}) that RRAT-like
periodicity was inconsistent with the bursts of at least one repeating FRB.

The fact that most distributions of intervals between bursts are bimodal
may cloud this picture.  If the shorter intervals, typically tens of ms,
represent substructure within bursts separated by (typically) a few minutes,
then it is not possible to define the stable near-exact period required by
RRAT-like periodicity.
\subsection{Periodic Modulation of Activity}
In a magnetar model of FRB the bursts themselves may be stochastic.
However, they cannot emit isotropically because a neutron star magnetic
field cannot be isotropic.  An observation has a detection threshold, and
the probability that a burst exceeds that threshold must depend on the
orientation of the field with respect to the observer.  That probability
varies with the rotation period (unless the field is azimuthally symmetric
about the rotation axis).  This model predicts that the rate of detected
bursts varies with the neutron star rotation period.  In this note we search
for, and set bounds on, such a periodic variation.
\section{Parameters and Phase Drift}
\citet{Z25} observed FRB 20240114A over eight months, which sets a lower
bound on its age.  No repeating FRB has ever been reported to turn off or
even to show a secular decrease in its activity, although their activity
varies greatly.  This suggests that ages of observed repeating FRB are
$\gtrsim 1\,$y, possibly $\gg 1\,$y.  FRB 20121102 has been observed to
be active for $> 10\,$y.  Here we assume FRB 20240114A had an age of 1 y
when observed on March 12, 2024; numerical results are readily scaled.

A magnetar with a dipole moment component perpendicular to its rotational
axis $\mu = \mu_{33} 10^{33}\,$Gauss-cm$^3$ (corresponding to an equatorial
field of about $10^{15}\,$Gauss), born spinning comparatively rapidly, has
an angular rotation frequency
\begin{equation}
\label{omega}
\omega = \sqrt{{3 \over 2}{I c^3 \over \mu^2 A}} \approx 36 I_{45}^{1/2}
\mu_{33}^{-1} A_y^{-1/2} \text{s}^{-1}
\end{equation}
at age A, where the moment of inertia $I = I_{45} 10^{45}\,$g-cm$^2$ and
$A_y$ is its age in years \cite{LGS}.  Eq.~\ref{omega} corresponds to a
rotational period 0.176$I_{45}^{-1/2}\mu_{33}A_y^{1/2}\,$s, and is
consistent with observed SGR periods 2--12 s for their known ages of a few
thousand years.
\section{Results}
In order to search for a periodic modulation of irregularly spaced events
we perform a periodogram \citep{S98,S06}.  There are few enough bursts and
few enough possible frequencies (integer multiples of $2\pi/T$, where $T$ is
the duration of the observation) up to $200\pi$Hz, to allow for periods as
short as 0.01 s ({\it cf.\/} Eq.~\ref{omega}).  This assumes that $\mu_{33}$
is not $\ll 1$, as for known Galactic magnetars.  It is plausible that a
very active magnetar-FRB must be strongly magnetized; see also the energetic
argument of \citet{Z25} pointing to large magnetic fields, although this
depends on uncertain assumptions about beaming and radiative efficiency.

The periodogram of events occurring at times $\{t_i\}$ is defined
\begin{equation}
\label{Pdefinition}
P(\omega) = \sqrt{\left(\sum_i \cos{(\omega t_i)}\right)^2+
\left(\sum_i \sin{(\omega t_i)}\right)^2}.
\end{equation}
It would be possible to weight the terms to account for the strength (flux
or fluence) of each burst, but then the small number of very strong bursts
would increase the stochastic noise.

\begin{figure}
\centering
\includegraphics[width=6.2in]{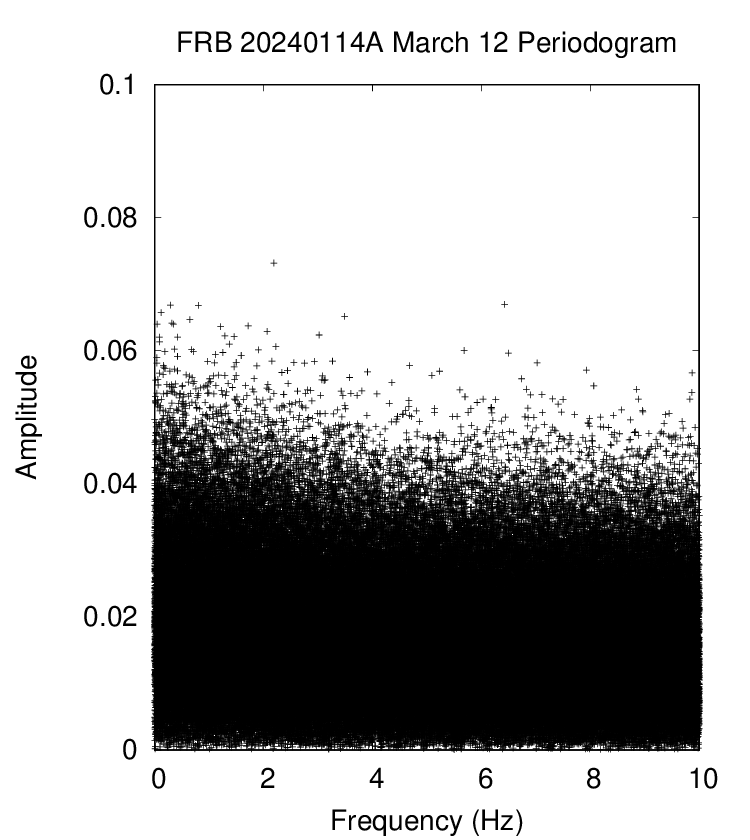}
\caption{\label{periodogram}{Unnormalized} periodogram of 3196 bursts of
FRB 20240114A in a 15628 s observation March 12, 2024 (MJD 60381), data from
\citet{Z25}.  Extension of the calculation to frequencies of 100 Hz
(implying a neutron star with $\mu_{33} \ll 1$ unless its age is $\ll 1\,$y)
produced no higher peaks than shown here.}
\end{figure}

Results for frequencies (in cycles/s, not radians) up to 10 Hz are shown in
Fig.~\ref{periodogram}.  Extending the search to frequencies up to 100 Hz
found no $P(\omega)$ larger than the largest found below 10 Hz (at a
frequency of 2.19 Hz).  The slight rise of the periodogram at the lowest
frequencies indicates a modest slow variation of the rate of detections
through the 15628 s (4.34 hours) of observation.  This rise may be
intrinsic, or it may be the consequence of a variation of instrumental
sensitivity with zenith angle.  {It may produce a similarly small
variation in the sensitivity of the search for periodicity as a function of
frequency.}
\section{Frequency Derivative}
For a rotor spinning down as a result of its dipole radiation
\begin{equation}
\label{omegadot}
{\dot \omega} = -{\omega \over 2 A} \approx -5.7 \times 10^{-7}\mu_{33}^{-1}\
A_y^{-3/2} \text{s}^{-2}.
\end{equation}

A large $|\dot\omega|$ blurs the signature of periodicity.  For an
observation of duration $T$ the phase drift of an oscillation from the
midpoint of the observation (to which it is fitted) {with an assumed
constant $\dot\omega$} to its ends is
\begin{equation}
\Delta\phi = {1 \over 8}{\dot \omega}T^2 \approx 17 \mu_{33}^{-1}
A_y^{-3/2}\ \text{radian},
\end{equation}
where for the observation on March 12, 2024 $T = 15628\,$s.  If $A_y > 10$,
$\Delta\phi < 0.55\mu_{33}^{-1}\,$radian and spindown during the observation
may have little effect on the detectability of periodicity, while if
$A_y < 5$, $\Delta\phi > 1.6\mu_{33}^{-1}\,$radian and detectability is
significantly reduced.  Of course, both $A$ and $\mu_{33}$ are unknown
{\it a priori.\/}  The duration over which the source has been observed sets
a absolute lower bound to $A$ {at its most recent observation.}  The
{statistical} improbability of discovering a repeating FRB very soon
after its birth and the radio opacity of very young SNR {imply that this
bound is conservative.  The absence of systematic long term trends in the
properties of a repeating FRB implies that $A$ is at least several times the
absolute lower bound (more than a decade for FRB 20121102A).  No such bounds
apply to apparent non-repeaters.}

A modified periodogram allows for steady frequency drift:
\begin{equation}
\begin{split}
C(\omega,{\dot\omega})&= \sum_i \cos{\left[(\omega t_i) + {1 \over 2}
{\dot\omega} (t_i-t_m)^2\right]}\\
S(\omega,{\dot\omega})&= \sum_i \sin{\left[(\omega t_i) + {1 \over 2}
{\dot\omega} (t_i-t_m)^2\right]}\\
P(\omega,{\dot\omega})&= \sqrt{C(\omega,{\dot\omega})^2
+S(\omega,{\dot\omega})^2},\\
\end{split}
\end{equation}
where $t_m$ is the midpoint of the data range.  We calculate this with
${\dot\omega} \in [-1 \times 10^{-6}, +1 \times 10^{-6}]\,$s$^{-2}$ in steps
of $10^{-8}$ s$^{-2}$.  Again, no significant peaks were found for any
$(\omega,{\dot\omega})$ with $2\pi/T \le \omega \le 200\pi/$s and the
preceding range of $\dot\omega$.  Yet greater $\omega$ would be inconsistent
with the magnetic moment implied by the magnetar hypothesis and an age
$> 1\,$y.
\section{Sensitivity}
In order to determine the sensitivity of the periodogram to periodic
variation of the rate of bursts we introduce such variation artificially.
{A burst at time $t$ is removed from the data with a probability
$M[1+\cos{(\omega_m t)}]$, where $M$ is the modulation half-amplitude and
$\omega_m$ its (angular) frequency.  We choose a frequency (in cycles) of
1 Hz; the weak dependence of the periodogram (Fig.~\ref{periodogram}) on
frequency shows that the sensitivity depends only weakly on frequency {a
stable periodic modulation would produce a large component of the
periodogram at its frequency, {which is not found}.}

When there are a large number of frequencies {in the periodogram a plot
like Fig.~\ref{periodogram} would be unclear.}  The clearest {signal of
periodicity would be an outlier in its distribution of} amplitudes. 
This would be clearer in the amplitude distribution than in the periodogram
itself which contains $1.5 \times 10^6$ points.
\cite{NK26} did this in a search for periodicity in the X-ray
emission of Sco X-1 in which there were $2^{29} \approx 5.37 \times 10^8$
frequencies and amplitudes, that could not be readily and legibly plotted.
Fig.~\ref{amplitudes} {shows the distribution of amplitudes for $M=0$
(the actual data) as in Fig.~\ref{periodogram} but extended to frequencies
up to 100 Hz} and with several larger values of imposed modulation $M$
at 1 Hz, a plausible spin frequency of a young magnetar.
Fig.~\ref{periodogram} shows that the noise level, and therefore sensitivity
to periodic variation, are nearly independent of the assumed period of
variation.
\begin{figure}
\centering
\includegraphics[width=6.2in]{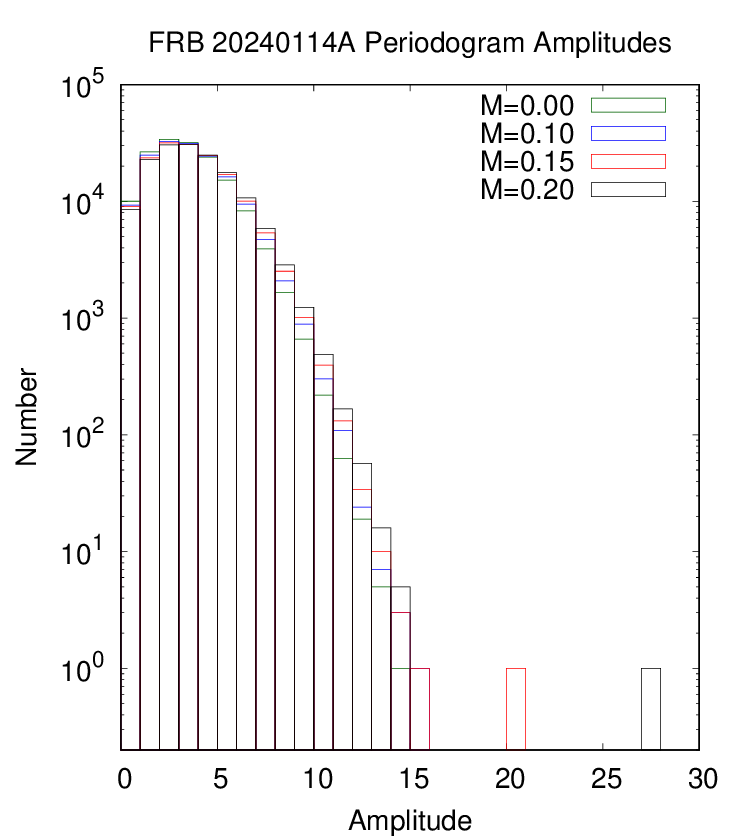}
\caption{\label{amplitudes}Distribution of amplitudes of the periodogram of
the March 12, 2024 observation \citep{Z25} of FRB2024014A, with stochastic
modulations of various amplitudes $M$ imposed.  {Amplitudes are defined
as the periodogram of Eq.~\ref{Pdefinition} normalized by the number of
bursts.}  A $M = 0.15$ {(15\% half-amplitude)} modulation of the event
rate is readily detected but a $M = 0.10$ {(10\% half-amplitude)}
modulation cannot be.  {Amplitudes are the periodogram
Eq.~\ref{Pdefinition} normalized by its mean.}  The bin width is one unit of
amplitude, chosen so that rare outliers are apparent.}
\end{figure}
\section{Periodicity?}
In the model of steady dipole spindown, Eq.~\ref{omegadot} shows that large
negative $\dot\omega$ only occurs for neutron stars that are both young and
comparatively weakly magnetized; in other words, not magnetars.
Quantitatively, the failure to find periodicity (for any possible $\omega$)
with ${\dot\omega} \in [-10^{-6},10^{-6}]$ s$^{-2}$ indicates $\mu_{33} \le
0.57 A_y^{-3/2}$ {if the modulation amplitude of the received signal $M
\ge 0.15$}.  If $A > 1\,$y, the implied $\mu_{33}$ would be below the
range considered to characterize magnetars.  Larger $|\dot\omega|$ outside
the range calculated would require $\omega$ so large as to be inconsistent
with Eq.~\ref{omega} and the assumptions of magnetar fields and $A > 1\,$y.

The possible presence of phase drift is an obstacle to coherent combination
of data from observing sessions separated by days.  Even if $\dot \omega$
were allowed for, over such longer intervals higher derivatives of $\omega$
become significant; measurements of radio pulsars show that these higher
derivatives may vary unpredictably.  %

\section{Conclusion}
The magnetar model of FRB makes no quantitative prediction of the magnitude
of modulation of the event rate at the magnetar's rotation frequency.  We
therefore cannot say that our upper bound of 15\% modulation in FRB
20240114A disproves this model.  The bound is much lower than the modulation
of radiated intensities of known magnetic neutron stars (radio pulsars,
accreting neutron star X-ray sources, magnetars in their quiescent X-ray
emitting state (Anomalous X-Ray Pulsars) and in their Soft Gamma Repeater
outbursts), that are generally ${\cal O}(50\%)$, but without a quantitative
theory of magnetar FRB any inference must be uncertain.

If the magnetar is not triaxial, magnetar models may still be consistent
with tight upper limits on rotational modulation of the rate of detected FRB
if either
\begin{enumerate}
\item The radiation pattern is azimuthally symmetric about the rotation
axis;
\item or, the rotation axis is aligned with the magnetic axis and the line
of sight, and emission is collimated along that axis.
\end{enumerate}
The former possibility might require the magnetic field to be nearly
azimuthally symmetric ($m=0$ components of all significant multipole orders
dominating), perhaps implausible.  The latter possibility has the advantage
that curvature radiation is aligned with the magnetic axis and may propagate
along field lines.  Very active repeating FRB may be those with this
favorable orientation, so that most of their activity is observable, while
apparently non-repeating FRB may be those with mis-aligned axes, so that we
only observe their bursts when their magnetic axes are, fortuitously but
rarely, aligned with our line of sight.


The suggestion that repeating FRB have rotational and magnetic axes aligned
with the line of sight, while apparent non-repeaters do not, implies that
we only observe a tiny fraction of the bursts of apparent non-repeaters
but also that we also observe a similarly tiny fraction of the repeaters.

Eqs.~\ref{omega} and \ref{omegadot} are inapplicable if the magnetar is
accreting.  Known magnetars (SGR) are all single objects, but there is no
compelling reason why some magnetars could not be born in and remain in
mass-transfer binaries; this is why our search included ${\dot\omega} > 0$.
Known mass-transfer binaries containing neutron stars radiate a thermal
X-ray spectrum, with no evidence of coherent radio emission (and usually no
evidence of particle acceleration at all, although microquasars like Sco X-1
are exceptions).  However, all these known accreting neutron stars have
magnetic fields below the magnetar range.  Accreting magnetars as a source
of FRB appear not to have been considered.
\section*{Data Availability}
This theoretical study produced no new data.
\section*{Acknowledgments}
I thank T. Piran for discussions and the Hebrew University, Jerusalem,
Israel for hospitality during the completion of this work.

\end{document}